\documentclass[twocolumn,english,prl,amssymb,aps,superscriptaddress,showpacs,twocolumn,amsmath,showkeys,floatfix,]{revtex4-1}
\usepackage[T1]{fontenc}
\usepackage[latin9]{inputenc}
\usepackage{geometry}
\usepackage[active]{srcltx}
\usepackage{amsmath}
\usepackage{graphicx}
\usepackage{esint}
\usepackage{epsfig}
\usepackage{epstopdf}
\usepackage{physics}
\usepackage{color}
\usepackage{natbib}
\usepackage{upgreek}
\usepackage{mathtools}
\usepackage{soul}
\usepackage{lipsum}
\usepackage[dvipsnames]{xcolor}
\usepackage{color}
\definecolor{blue}{rgb}{0,0,1}
\definecolor{green}{rgb}{0,1,0}
\definecolor{purple}{rgb}{0.5,0,1}
\makeatletter
\usepackage{babel}

\makeatother

\usepackage{babel}
\begin{document}

\title{Spin noise spectroscopy of optical light shifts}

\author{J. Delpy}
\affiliation{Universit\'e Paris-Saclay, CNRS, Ecole Normale Sup\'erieure Paris-Saclay, CentraleSup\'elec, LuMIn, Orsay, France}
\author{S. Liu}
\affiliation{Universit\'e Paris-Saclay, CNRS, Ecole Normale Sup\'erieure Paris-Saclay, CentraleSup\'elec, LuMIn, Orsay, France}\affiliation{East China Normal University, State Key Laboratory of Precision Spectroscopy, Shanghai, China}
\author{P. Neveu}
\affiliation{Universit\'e Paris-Saclay, CNRS, Ecole Normale Sup\'erieure Paris-Saclay, CentraleSup\'elec, LuMIn, Orsay, France}
\author{E Wu}
\affiliation{East China Normal University, State Key Laboratory of Precision Spectroscopy, Shanghai, China}
\author{F. Bretenaker}
\affiliation{Universit\'e Paris-Saclay, CNRS, Ecole Normale Sup\'erieure Paris-Saclay, CentraleSup\'elec, LuMIn, Orsay, France}
\author{F. Goldfarb}
\affiliation{Universit\'e Paris-Saclay, CNRS, Ecole Normale Sup\'erieure Paris-Saclay, CentraleSup\'elec, LuMIn, Orsay, France}
\affiliation{Institut Universitaire de France (IUF)}

\begin{abstract}

Light induced non-equilibrium spin noise spectroscopy is theoretically and experimentally shown to be an efficient technique to reveal the structure and the coherent effects in the probed transition. Indeed, using metastable helium, the spin noise spectrum is shown to exhibit a dual-peak structure around the Larmor frequency. This previously unobserved feature is due to the light shifts of the involved levels and strongly depends on the probe power, detuning, and polarization orientation. Both numerical and analytical models reproduce very well the details of the split spin noise spectra: this technique thus allows a simple and direct measurement of the light shifts, and its polarization dependence permits to reveal the level structure in a non ambiguous manner. 

\end{abstract}

\maketitle

\section{Introduction}

Over the last few decades, with the broad availability of real-time spectrum analysers \cite{crooker2004spectroscopy}, spin noise spectroscopy has proved to be an efficient method to probe the stochastic fluctuations of magnetization in a strictly non-polarized atomic ensemble \cite{sinitsyn2016theory}. By using a non-perturbative, off-resonant linearly polarized probe light and a DC transverse magnetic field \cite{aleksandrov1981}, spin fluctuations of the system are converted into modulated fluctuations of the orientation (as depicted in Fig.\,\ref{setupandresults}(a)) or the ellipticity of the probe polarization. Measuring these fluctuations provides diverse information regarding the atomic system and the interactions between the spins and their environment, like their relaxation time at thermal equilibrium \cite{sinitsyn2016theory,muller2008spin}. First demonstrated in atomic vapors, this method was quickly adapted to strongly-correlated structures, like semi-conductors \cite{oestreich2005spin, muller2008spin}, quantum wells \cite{poltavtsev2014spin} or quantum dots \cite{CrookerQuantumDots}. 

In spite of all the information that can be obtained at thermal equilibrium, the spin dynamics is necessarily constrained by the fluctuation-dissipation relations \cite{Kubo_1966}. For this reason, spin noise spectroscopy has been extended further to non-equilibrium conditions. \textcolor{black}{For example, Refs. \cite{Yue_Evolution, Nonequilibrium_sinitsyn} focus on the theory of the non-equilibrium regime induced in condensed matter structures by AC and DC electric and magnetic fields, while Refs. \cite{NonEqQuantumDot1, NonEqQuantumDot2} report experimental observations in quantum dots. In atomic vapors, such a regime, although less discussed, has also raised a lot of interest, e.g., to measure more precisely the structure of the involved transition, or to investigate the response of the system to one or several resonant fields that drive coherences between the  states. Moreover, Ref. \cite{swar2018measurements} reports SNS in strong optical pumping conditions, while} the work of Glasenapp \textit{et al} \cite{glasenapp2014spin} in 2014 shows the interest of inducing coherences between Zeeman sub-levels with the help of off-resonant and resonant AC magnetic fields. This method provides various details on the ground level structure at the price of a more complex experimental setup.

In this paper, we thus wonder whether the off resonant probe light field itself can be used to create a non-equilibrium regime in the atomic system, without any modification of the experimental setup. The question is to know whether the laser field can play both the role of the probe -- experiencing the fluctuating Faraday rotation and ellipticity induced by the spin noise -- and the role of a driving field that induces coherences between different atomic states, even if it is far from resonance. To answer these questions, we choose to perform spin noise-spectroscopy in a metastable helium vapor cell, in the vicinity of the $2^3S_1 \leftrightarrow 2^3P_0$ transition, which is very well isolated from other transitions and thus relatively easy to model. 
We thus hope that, using this very simple level scheme, the presence of the probe field will allow new features revealing the coherences to appear.

We first introduce the experimental setup and show how these unusual features appear in spin noise spectra. We then explore the influence of various experimental parameters such as the probe power, detuning, and polarization orientation. We compare these behaviors with numerical computations based on transit-induced spin simulations. We finally discuss the physical origin of the observed splittings in the spectra, by using an analytical model based on the relations between the spin dynamics and the eigenfrequencies of the system.

\section{Experimental setup }

\begin{figure}
    \centering
    \includegraphics[width=\columnwidth]{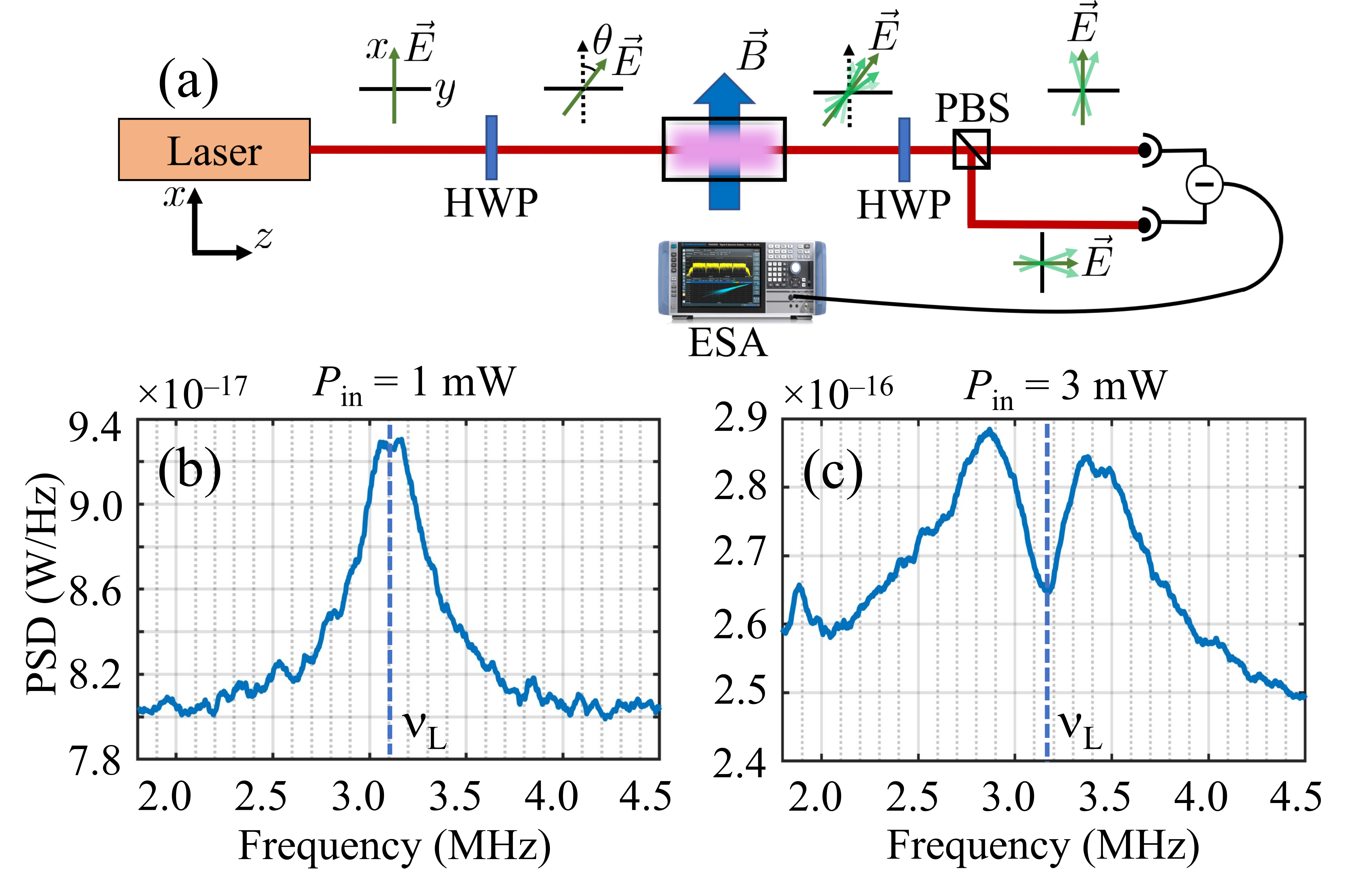}
    \caption{(a) Experimental setup used. Spin fluctuations are converted into polarization orientation and ellipticity fluctuations, modulated at the Larmor frequency and recorded using a balanced detection. (b) Faraday rotation noise spectrum obtained for a detuning $\Delta/2\pi=1.5 \, \mathrm{GHz}$, an input power $P_{\mathrm{in}}=1\, \mathrm{mW}$, and an input polarization direction $\theta=0^{\circ}$. (c) Same as (b) for $P_{\mathrm{in}}=3\, \mathrm{mW}$. Two clearly distinct sub-resonances appear on the spectrum. For (b) and (c), the resolution bandwidth is 27\,kHz.}
    \label{setupandresults}
\end{figure}

The experimental setup, which  is described in more details in Ref. \cite{ShikangAccepte},  is depicted in Fig.\,\ref{setupandresults}(a). The probed atomic sample  is a vapor cell filled with 1 Torr of $^4\mathrm{He}$ at room temperature. Using a RF discharge, a fraction of the atoms are excited from the $2^3S_0$ fundamental level to the metastable $2^3S_1$ level. This level is composed of three Zeeman sublevels, separated by a Larmor frequency $\nu_L$ in the MHz range by the application of a transverse DC magnetic field of a few Gauss.
The input probe light, propagating along $z$, is linearly polarized at an adjustable angle $\theta$ with respect to the direction $x$ of this magnetic field.

As the $^4$He atom does not exhibit any nuclear spin, the $2^3S_1 \leftrightarrow 2^3P$ transition has only a fine structure. The three upper levels, namely the states $\ket{2^3P_{0,1,2}}$, are separated by nearly 30 GHz and 2.3 GHz, respectively. Since the Doppler width in our experimental conditions is about 0.8 GHz \cite{goldfarb2008observation}, we choose to probe the $2^3S_1 \leftrightarrow 2^3P_0$ (D$_0$) transition only using  blue-detuned probe light and neglecting the effect of the two other transitions.

At the output of the cell, the laser travels through a half-wave plate (HWP) and a polarization beam splitter whose outputs are sent to a balanced detection composed of two identical photodiodes. Adjustment of the orientation of the HWP allows us to balance the detection and  record the fluctuations of orientation of the polarization (see \cite{ShikangAccepte}). 
Adding a quarter-wave plate before the half-wave plate also allows to measure the ellipticity fluctuations by converting them into orientation fluctuations.
The photocurrent difference at the output of the balanced detection is sent to an electronic spectrum analyser (ESA), which provides us with its power spectral density. Although this experimental setup is common to the vast majority of gas cells spin noise experiment, we report here the observation of very uncommon noise spectra when the probe power is increased.

\section{Experimental results}
By using a low-power, far-detuned laser beam, detecting the fluctuating Faraday rotation of the probe light corresponds to recording 
the fluctuations of the circular birefrigence of the vapor. This noise is directly due to the fluctuations of the populations of the Zeeman sub-levels  \cite{crooker2004spectroscopy}. These populations oscillate periodically under the action of the transverse magnetic field, which translates the zero-frequency spin noise to the Larmor frequency $\nu_{\mathrm{L}}$ of the system \cite{sinitsyn2016theory}. The same effect holds for the ellipticity fluctuations recorded when the quarter-wave plate is inserted, however the spin arrangement responsible for the noisy linear birefringence of the sample exhibits oscillations both at the Larmor and twice the Larmor frequency \cite{fomin2020spin,ShikangAccepte}. Either way, usually only one isolated Lorentzian-shaped resonance is visible at each frequency \cite{sinitsyn2016theory}. Figure \ref{setupandresults}(b) reproduces a typical Faraday rotation noise spectrum obtained far from resonance, when the laser power is set to 1~mW with a beam diameter of 0.6 mm. However, in the same conditions but increasing the probe power up to 3~mW, one obtains the spectrum of Fig.\,\ref{setupandresults}(c). Surprisingly, the Faraday rotation noise spectrum is split into two components around the Larmor frequency $\nu_{\mathrm{L}}$. The two sub-resonances are symmetrical with respect to $\nu_{\mathrm{L}}$. The same effect is observed for the ellipticity noise spectrum.
\paragraph{}
\begin{figure}
    \centering
    \includegraphics[width=\columnwidth]{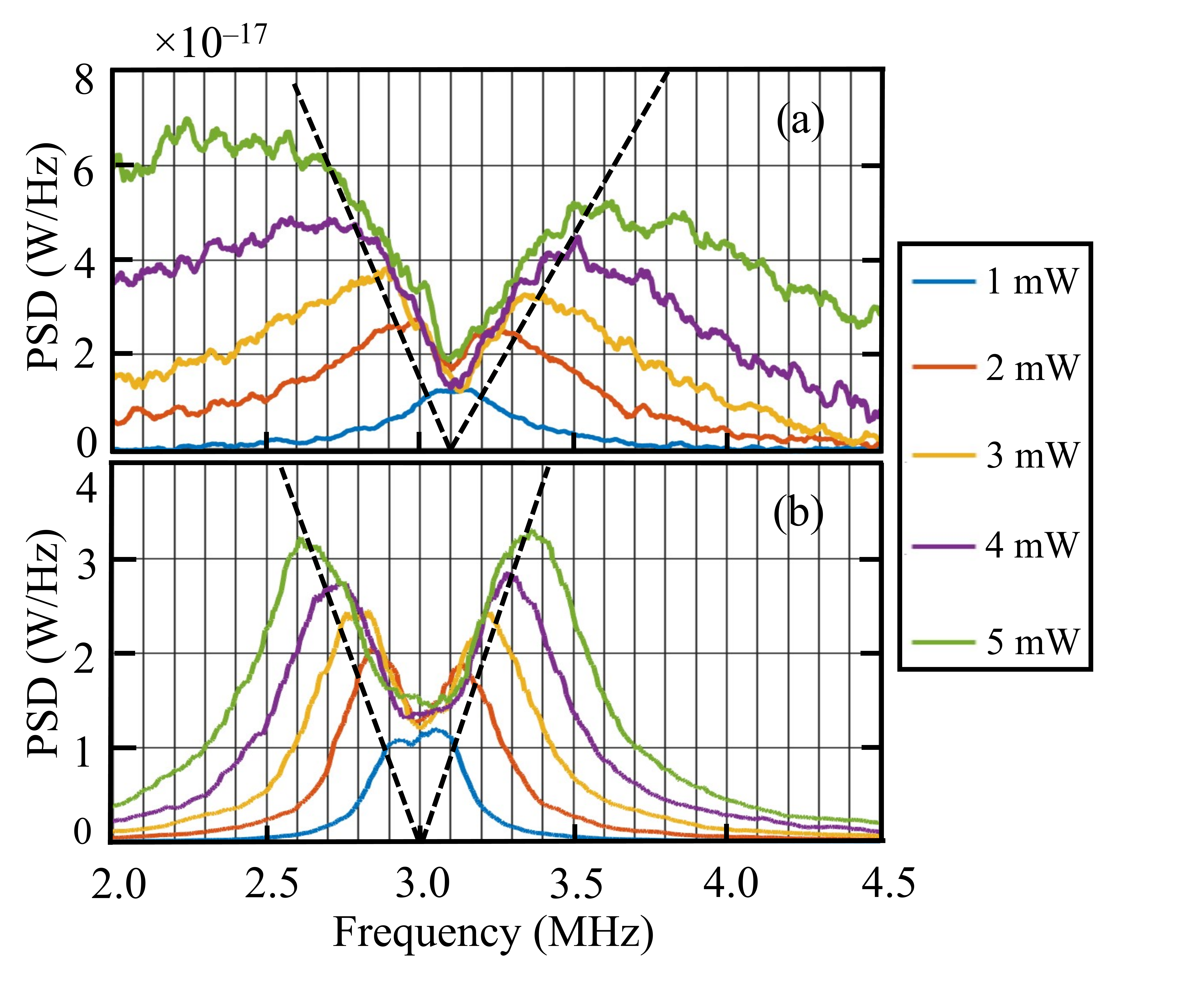}
    \caption{(a) Experimental Faraday rotation noise spectra obtained 1.5~GHz from resonance, with the probe light polarization aligned with the DC transverse magnetic field ($\theta=0^{\circ}$). The probe power varies from 1 to 5~mW. (b) Numerical simulations obtained with $\nu_L = 3\, \mathrm{MHz}$, and a Raby frequency varying from 40 to 90 MHz, corresponding to $P_{\mathrm{in}}=1$ to 5 mW.}
    \label{power_analysis}
\end{figure}

Figure\,\ref{power_analysis}(a) shows several Faraday rotation noise spectra obtained with a laser blue detuned by $\Delta/2\pi = 1.5\,\mathrm{GHz}$  from the center of the D$_0$ transitions. The Larmor frequency is set at 3.1~MHz, and the magnetic field and the probe polarization directions are aligned ($\theta=0^{\circ}$). The power of the probe beam is progressively increased from 1~mW to 5~mW. 
The noise floor corresponding to the shot noise  is substracted. One can clearly see the gradual splitting of the rotation noise peak into two resonances, which are perfectly visible when the probe power reaches 2~mW or more. This splitting seems to increase linearly with the laser power. Moreover, one also observes a broadening of the sub-resonances, which we attribute to the residual excitation of the atoms from the lower to the upper level, thus shortening the effective lifetime of the $2^3S_1$ Zeeman sub-levels. 

These spin noise spectra could be theoretically reproduced by numerically simulating the model presented in details in Ref.\,\cite{ShikangAccepte}. This model is based on the evolution of the density matrix of the system under the fluctuations of the atomic levels populations. The response of the atoms is added to the probe field using the thin-cell approximation. Fluctuations of the Zeeman sub-level populations are numerically introduced to mimic the transit of the atoms through the beam. The results of such simulations, performed using the parameters extracted from the experiment, are presented in Fig.\,\ref{power_analysis}(b): the dual-peak shape of the Faraday rotation noise as well as the values of the frequency splitting and the noise maxima are well reproduced. Only the peak width is underestimated, since our model does not take velocity changing collisions into account \cite{Toto} so that the saturation of the transition is not perfectly simulated.

The data of Figs.\,\ref{setupandresults} and \ref{power_analysis} were obtained for a fixed direction of the input probe linear polarization direction parallel to the magnetic field ($\theta=0^{\circ}$). Further insight can be gained by investigating the dependence of the spin noise spectrum on the input polarization direction.

\begin{figure}
    \centering
    \includegraphics[width=0.97\columnwidth]{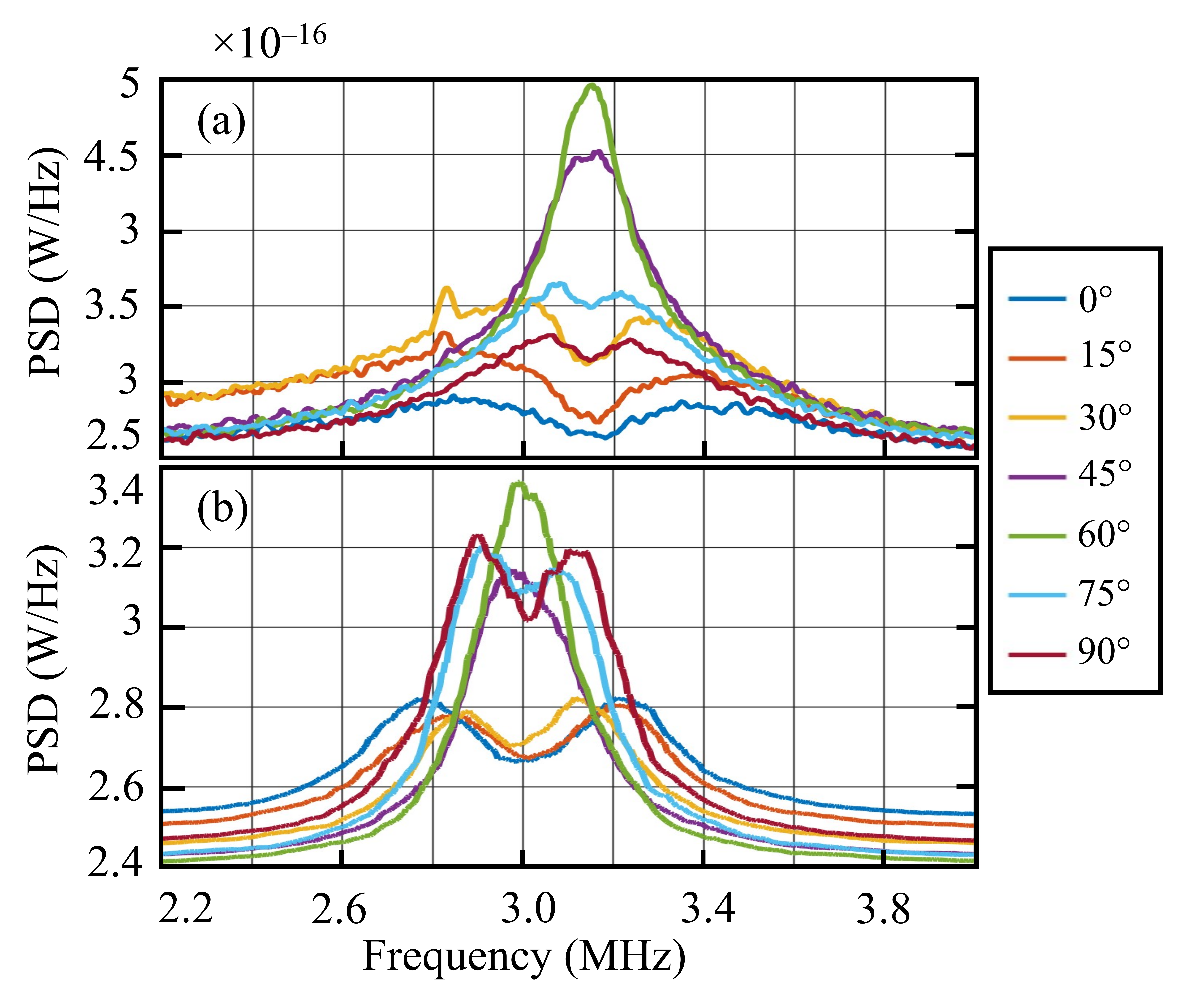}
    \caption{(a) Experimental Faraday rotation noise spectra obtained for a laser power P$_{\mathrm{in}} = 3$mW, $\Delta/2\pi =1.5$ GHz, and $\theta$ varying from 0 to 90$^{\circ}$. ESA parameters : RBW~=~30 kHz, VBW~=~5.1~Hz. 
    (b) Simulated spectra using a Rabi frequency $\Omega/2\pi = 70 \, \mathrm{MHz}$.}
    \label{polar_dependance}
\end{figure}
\paragraph{}
Figure \ref{polar_dependance}(a) shows the experimental Faraday rotation spectra obtained with the probe power set to 3\,mW as before, for $\Delta/2\pi = 1.5$ GHz, and for an input probe polarization direction varying from $\theta=0^{\circ}$ to $\theta=90^{\circ}$.
The separation between the two noise peaks strongly depends on the angle $\theta$ : the splitting is maximum for $\theta=0^{\circ}$, and decreases until it completely disappears in the vicinity of $\theta = 60 ^{\circ}$. Between 60 and 90$^{\circ}$, the splitting exhibits a revival, but remains smaller than for small $\theta$ angles. The maxima of the different spectra also strongly depend on the polarization direction, which we attribute to the polarization-dependent saturation of the transitions (see \cite{ShikangAccepte} for details). The corresponding simulation results are shown in Fig.\,\ref{polar_dependance}(b). The Rabi frequency is chosen to be 70 MHz, corresponding to a 3 mW power beam. Here again, the simulation results reproduce the experimental polarization dependence : the splitting is maximum for $\theta = 0 ^{\circ}$, completely disappears near $\theta = 60^{\circ}$, and appears again between 60 and 90$^{\circ}$. The discrepancy between theory and experiments concerning the amplitudes of the resonances is again due to the lack of complete description of the saturation. 

\section{Physical interpretation}

The frequencies at which the noise resonances appear in the spectra can be related to the differences in energy between the Zeeman sub-levels of the lower level of the transition \cite{sinitsyn2016theory}. Some physical insight into these unusual observations can thus be gained by considering the influence of the light on the eigenenergies of the system. Despite the influence of the probe field being commonly overlooked in the literature, our observation of dual-peak noise spectra proves that it cannot be ignored, even far from resonance. Moreover, we show that this non-equilibrium regime carries interesting details on the lower state structure.

The evolution of the system is driven by the magnetic field oriented along $x$. 
We call $\ket{-1}_x$, $\ket{0}_x$, and $\ket{1}_x$ the Zeeman sublevels, eigenvectors of the spin operator along $x$, and $E_{-1}$, $E_0$, $E_1$ their energies. The dynamics of the system thus occurs at frequencies given by the differences between these energies, namely:
\begin{equation}
\left\{
\begin{array}{c c}
     \nu_+ =& \dfrac{E_{+1} - E_0}{h}\ ,\\
    \nu_- =&\dfrac{E_0 - E_{-1}}{h}\ .
\end{array}
\right.
\label{Eq04}
\end{equation}
In usual spin noise spectroscopy, these two frequencies are degenerate and equal to the Larmor frequency $\nu_{\mathrm{L}}$.

Now, the polarization of the probe light incident along the $z$ direction is sensitive to the difference of populations between the the Zeeman sublevels $\ket{-1}_z$ and $\ket{+1}_z$ obtained with the quantization axis along $z$. This population difference, and thus the Faraday rotation noise, oscillates at frequencies $\nu_+$ and $\nu_-$. The fact that the spectra of Figs.\,\ref{power_analysis} and \ref{polar_dependance} contain two peaks indicates that these two frequencies become different. This effect originates from the light shifts induced by the probe light on the Zeeman sublevels, as is now going to be explained in details.

Some off-resonant light impinging on an atomic sample coherently modifies the energy levels of the probed transitions \cite{cohen2012processus}, a phenomenon known as the AC Stark shift, or light shift, of the atomic states. In a two-level  system, it can be shown that this energy shift of the lower level reads, in the approximation $\Delta \gg \Omega, \Gamma/2, $ \cite{cohen2012processus,grynberg_aspect_fabre_cohen-tannoudji_2010}:
\begin{equation}
\hbar \delta = \hbar \dfrac{\Omega ^2}{4\Delta}\ .
    \label{Eq05}
\end{equation}

\begin{figure}
    \centering
    \includegraphics[width=\columnwidth]{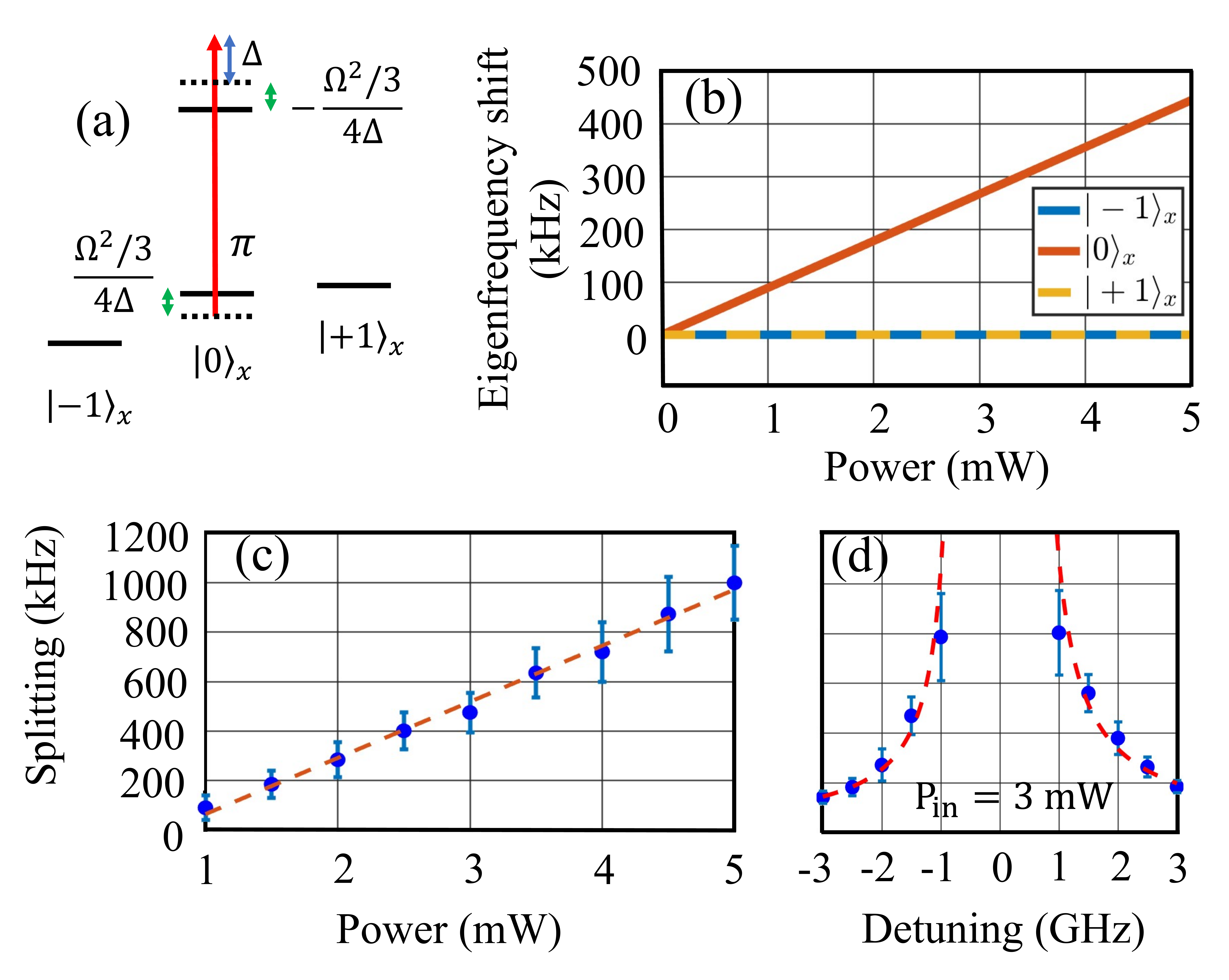}
    \caption{(a) Scheme of the energy levels and light shift experienced for $\theta = 0^{\circ}$. (b) Numerical calculations of the shift experienced by the three lower eigenfrequencies when increasing power, with $\theta = 0^{\circ}$, $\Delta/2\pi = 1.5 \, \mathrm{GHz}$.  (c) Blue dots : Experimental measurement of the frequency splitting in the same experimental conditions. Dashed orange line : linear fit of the datas. (d) Blue dots : Experimental measurement when varying detuning, with $P_{in} = 3 \, \mathrm{mW}$. Red dashed line : fit obtained with a hyperbolic function.}
    \label{power_detuning_study}
\end{figure}

Additionally, the upper level is shifted by the opposite amount $-\hbar \delta$. Such AC Stark shifts are usually measured using non-linear experiments such as pump-probe or multiphoton methods \cite{delsart:jpa-00209038, PhysRevA.93.023433, Palani_2021, CSStarkShift}. In the case of the $J=1\rightarrow J=0$ transition of our spin-1 system, the shift experienced by the lower Zeeman sub-levels depends on the Rabi frequency along each transition (see Fig. \ref{power_detuning_study}(a)). Figure \ref{power_detuning_study}(b) shows the computed frequency shifts $\delta/2\pi$ experienced by the $2^3S_1$ state sub-levels as a function of the probe power, when $\theta = 0^{\circ}$, i.e. only the $m=0\rightarrow m=0$ transition ($\pi$ transition) is excited, like in Fig. \ref{power_detuning_study}(a). In this case only the $\ket{0}_x$ ground state Zeeman sublevel is shifted by $\dfrac 1 3 \dfrac{\Omega^2}{4\Delta}$. The factor 1/3 results from the square of the corresponding Clebsch-Gordan coefficient and is the same for the three transitions. Consequently, the degeneracy between $\nu_- = \nu_{\mathrm{L}} + \delta/2\pi$ and $\nu_+ = \nu_{\mathrm{L}} - \delta/2\pi$ is lifted, and these frequencies appear in the Faraday rotation noise spectra. The AC-Stark shift thus becomes accessible using our spin noise spectroscopy setup.
As seen in Fig.\,\ref{power_detuning_study}(b), the theoretical splitting, given by $2\delta/2 \pi$, increases linearly with the laser power with a slope of  $180\,\mathrm{kHz/mW}$. This expectation is experimentally verified : Fig.\,\ref{power_detuning_study}(c) shows that the acquired resonances split up at a rate of $220 \mathrm{kHz/mW}$. \textcolor{black}{Experimental splittings were measured as the frequency difference between the peak maxima obtained by fitting the data with an empirical function. We  assume that the hole in between the resonances has a Lorentzian lineshape, which is well verified experimentally. Error bars are based on the uncertainty in the determination of the maxima.} Moreover, Fig.\ref{power_detuning_study}(d) reproduces the measured evolution of the splitting with the detuning for a 3 mW probe power. These data can be well fitted by a hyperbolic function (dashed red line), which confirms that the splitting scales like a light shift with respect to the detuning. \textcolor{black}{This experimental demonstration permits to reinterpret the theoretical predictions of Ref. \cite{NonLinearSNS} in terms of light shifts.}

\begin{figure}
    \centering
    \includegraphics[width=0.95\columnwidth]{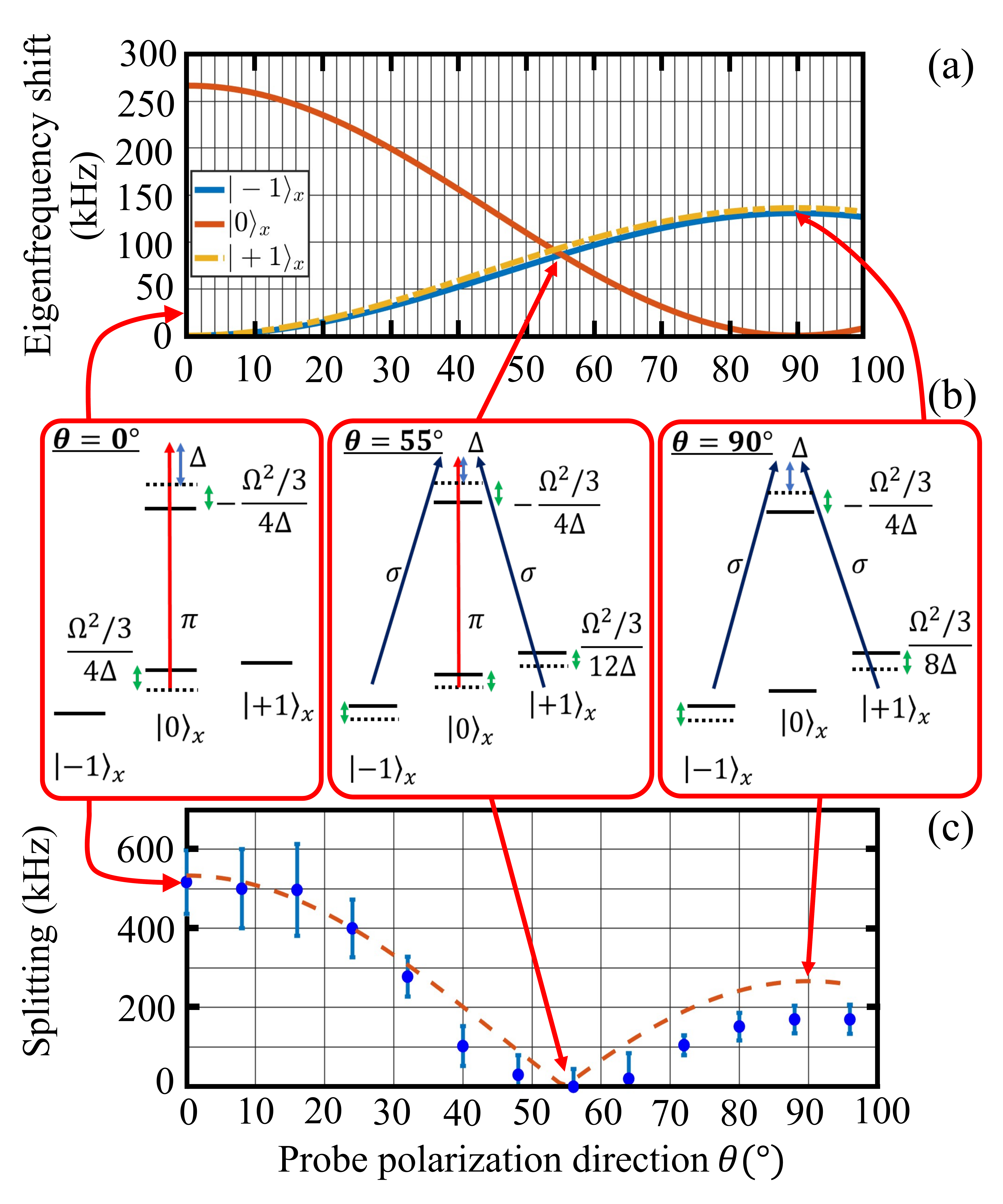}
    \caption{(a) Numerical calculations of the shift experienced by the three lower eigenfrequencies when varying the probe polarization direction, with $\Omega/2\pi = 70 \, \mathrm{MHz}$, $\Delta/2\pi = 1.5 \, \mathrm{GHz}$. (b) Light shift experienced in the three limit cases where only $\pi$ transitions are allowed, only $\sigma$ or both with equal intensities. (c) Comparison between experimental splitting (blue dots) and numerical calculations (dashed orange line) based on the eigenfrequencies of the system. }
    \label{polar_study}
\end{figure}

The polarization behavior of the splitting observed and simulated in Fig.\,\ref{polar_dependance} can be  explained considering the projection of the probe Rabi frequency on the three transitions, which depends on $\theta$. Figure \ref{polar_study}(a) shows the resulting dependence of the light shifts of the three Zeeman sublevels on the probe polarization direction $\theta$ for a fixed Rabi frequency equal to $70 \, \mathrm{MHz}$. When $\theta=0^{\circ}$ (see Fig.\,\ref{polar_study}(b)), as discussed above, only $E_0$ is modified, leading to a maximum splitting between the resonances. When $\theta$ increases, the projection of the Rabi frequency on the $\sigma^+$ and $\sigma^-$ transitions increases while the projection on the $\pi$ transition decreases. This leads to an increase of $E_{-1}$ and $E_{+1}$ and a decrease of $E_0$, thus reducing the gap between the resonances. At $\theta=55^{\circ}$ (middle of Fig.\,\ref{polar_study}(b)), the three sublevels experience the same light shift and $\nu_+=\nu_-$. Finally, for $55^{\circ}<\theta<90^{\circ}$, the Rabi frequency for the $\sigma$ transitions exceeds the one of the $\pi$ transition and the positions of the resonances are reversed. For $\theta = 90 ^{\circ}$, only the $\sigma$ transitions are excited and only the states $\ket{\pm1}_x$ see their energies displaced by the same quantity $\dfrac \hbar 3 \dfrac{\Omega^2}{8\Delta}$ at first order in $\Omega/\Delta$.

This polarization dependence in the light shift is a genuine signature of a spin-1 system, with only one sub-level for the upper state. Any other spin value or transition structure would lead to a different polarization dependence. This proves that the non-equilibrium regime in which the probe field drives coherences between the lower and upper levels is an effective technique to probe the structure of the transition.

\section{Conclusion}

We observed a splitting in the spin noise spectra of a metastable Helium vapor when using a few mW probe beam. The frequency difference between the two sub-resonances has been shown to increase linearly with the probe power, and to be inversely proportional to the detuning from resonance. These observations are well reproduced by our numerical model, which simulates the evolution of the density matrix of the system under the assumption of transit-induced spin fluctuations.

These observations have been successfully interpreted in terms of AC Stark shift experienced by the lower Zeeman sublevels under coherent driving of the light probe. We proved that this shift modifies their relative energy differences, and thus lifts the degeneracy between the two frequencies at which the populations of the states $\ket{-1}_z$, $\ket{0}_z$, $\ket{+1}_z$ oscillate. This non-equilibrium regime makes this spin noise spectroscopy a useful technique to measure optical light shifts.

Moreover, a non-trivial polarization dependence of the spin noise spectrum splitting was observed. Indeed, the resonance splitting evolves in a non monotonous manner when the light linear polarization direction is rotated apart from the magnetic field direction. This very specific evolution was proved to be a signature of a spin-1 system with a single excited state.
This polarization resolved study is thus a powerful tool for detailed investigation of the level structure. 

Further investigations include a more precise study the shape of the dual-peak resonances, which implies to better model the spin correlator under the action of the probe light field. It should also be interesting to look at higher spin systems, which can also lead to features specific to their level structure and spin oscillation modes.

\selectlanguage{english}

\begin{acknowledgments}
The authors acknowledge funding by the Institut Universitaire de France and the Labex PALM.
\end{acknowledgments}


%

\appendix

\end{document}